# Application of the coupled classical oscillators model to the Fano resonance build-up in a plasmonic nanosystem

P.A. Golovinski[1,2], A.V. Yakovets[1], E.S. Khramov[1]

[1]Moscow Institute of Physics and Technology (State University)
[2]Voronezh State Technical University

We study the excitation dynamics of Fano resonance within the classical model framework of two linear coupled oscillators. An exact solution for the model with a damped harmonic force is obtained. The details of growth a Fano profile under the harmonic excitation is shown. For incident ultra-wideband pulse, the reaction of a system becomes universal and coincides with the time-dependent response function. The results of numerical calculations clarify two alternative ways for experimental measurement of the complete characteristics of a system: direct observation of the system response to a monochromatic force by frequency scanning or recording of time-dependent response to the $\delta$-pulse. As a specific example, time-dependent excitation in a system consisting of a quantum dot and a metal nanoparticle is calculated. Then it is shown the applicability of the extended model of damped oscillators with radiative correction to describe the build-up a plasmon Fano resonance in a scattering of a femtosecond laser pulse by nanoantenna.

## 1. Introduction

Excitation of isolated energy level by a laser pulse is a well-known problem of quantum electronics, and its solution for long pulses with a great number of oscillations was described in detail [1, 2]. When the few-cycle pulses used, there occur specific changes in the behavior of two-level system that manifest themselves in violation monotonicity of the time dependence the transition probability caused by a weak field and in ability to control the level population with the use of the chirped pulses [3-5]. In a number of physical systems, an excited level is not isolated, and interaction of a discrete state with other degrees of freedom can be described in the form of interaction with a band structure. This is what autoionization states [6, 7], excited states of an atomic nucleus against the background of a continuous spectrum [8,9], molecular electron excitations interacting with vibration spectrum [10-12], or impurity states of a solid interacting with a matrix spectrum [13] look like. Resonances occurring in excitation of such type of states have a characteristic asymmetric shape due to interference of level and a band structure. The theory of resonances for a discrete state interacting with a continuum was developed independently by Fano and Feshbach, and resonances themselves that are manifested in a great diversity of physical systems are generally referred to as Fano-Feshbach resonances [14].

The development of the nanoscale systems physics and technology of has revealed the presence in these structures various new Fano-Feshbach resonances [15], which we will henceforth, in accordance with generally accepted practice, be called the Fano resonances. From the viewpoint of the development prospects for nanooptical and nanoelectronic devices, of particular interest are resonances caused by interaction of states in quantum dots with acoustical [16, 17] and optical [18-20] phonons as well as with plasmons in metal nanoparticles and nanoconductors [21-30]. The action of a strong resonance laser field has a significant effect on the structure of Fano resonances [31-34], changing both the parameters and the dependence shape itself. If an external interaction is sufficiently weak, the system response becomes linear with respect to the field, so the change in average values of so-named coordinate operators is described by the classical equations of motion for coupled oscillators [35]. We consider the model of two coupled classical oscillators with time-dependent force to demonstrate the dynamic of formation a Fano profile.

**2. Classical model**

Fano resonance is a universal phenomenon since the manifestation of destructive interference does not depend on the nature of a medium. The importance of Fano resonances for nanophysics consists in information they contain that concerns the configuration of interacting modes and internal potential fields in low-dimension structures. This information can be obtained from the wave interference pattern in different channels. The classical model of Fano resonances can be constructed in the form of two weakly coupled oscillators excited by an external force [35-37], so a Fano resonance can be rather easily simulated with the use of equivalent electric circuits [38]. Following [36], we consider the classical equations of motion for two coupled oscillators

$$\ddot{x}_1 + \gamma_1 \dot{x}_1 + \omega_1^2 x_1 + v x_2 = f(t), \qquad (1)$$

$$\ddot{x}_2 + \gamma_2 \dot{x}_2 + \omega_2^2 x_2 + v x_1 = 0,$$

where $v$ describes elastic coupling of oscillators. We seek the response of a system with zero boundary conditions to an arbitrary field $f(t)$ by the use of the Laplace transform. In the transform domain, the system of equations (1) can be written as

$$s^2 X_1 + \gamma_1 s X_1 + \omega_1^2 X_1 + v X_2 = F(s), \qquad (2)$$

$$s^2 X_2 + \gamma_2 s X_2 + \omega_2^2 X_2 + v X_1 = 0.$$

Then the solutions for transforms are of the form

$$X_1(s) = \frac{(s^2 + \omega_2^2 + \gamma_2 s) F(s)}{(s^2 + \omega_1^2 + \gamma_1 s)(s^2 + \omega_2^2 + \gamma_2 s) - v^2}, \tag{3}$$

$$X_2(s) = -\frac{v F(s)}{(s^2 + \omega_1^2 + \gamma_1 s)(s^2 + \omega_2^2 + \gamma_2 s) - v^2}.$$

The response to the harmonic force is obtained from equation (3) by the substitution $s = i\omega, F(s) = 1$ that means a coming to the equivalent Fourier representation.

The response to an arbitrary pulse can be found by the inverse Laplace transform, in which the contribution of individual oscillating components is defined by pole residues. The final solutions in the time domain have the form

$$x_j(t) = \sum_n \frac{P_j(s_n)}{Q'(s_n)} e^{s_n t}. \tag{4}$$

Here $s_n$ are the values of the roots of the polynomial $Q(s)$ in the denominators of the equations (3), $Q'(s)$ is the derivative of $Q(s)$, $P_j(s)$ are the functions in the numerators of the equations (3). For a limit of big time, oscillations in a system are vanished for any pulse of finite duration. For the case, when harmonic component of the external force is survived, the stationary oscillations of constant amplitude are set. Let us consider the system response to an external force in the form of damped oscillations

$$f(t) = f_0 e^{-\lambda t} \sin(\omega t), t > 0 \tag{5}$$

that has the Laplace transform

$$F(s) = \frac{\omega}{(s+\lambda)^2 + \omega^2}. \tag{6}$$

To find the inverse Laplace transform, we determine the poles in the equation (3) in view of a chosen pulse shape. The poles of $F(s)$ are $s_{1,2} = -\lambda \pm i\omega$, and we will find the poles of $Q(s) = (s^2 + \omega_1^2 + \gamma_1 s)(s^2 + \omega_2^2 + \gamma_2 s) - v^2$ under the condition of smallness of the oscillator coupling parameter: $\omega_2^2 - \omega_1^2 \gg v$. The unperturbed roots at $v = 0$ are $s_n = -\gamma_j/2 \pm i\sqrt{\omega_j^2 - \gamma_j^2/4}, n = 3,4,5,6$. Corrections $\delta s_n$ due to interaction of the oscillators are found from the equations

$$\delta s_{3,4} = \frac{v^2}{(2s_{3,4} + \gamma_1)(s_{3,4}^2 + \omega_2^2 + \gamma_2 s_{3,4})}, \tag{7}$$

$$\delta s_{5,6} = \frac{v^2}{\left(2s_{5,6} + \gamma_2\right)\left(s_{5,6}^2 + \omega_1^2 + \gamma_1 s_{5,6}\right)}.$$

To make specific calculations, we have taken the parameters [36] of the model system as $\omega_1 = 1$ eV, $\omega_2 = 1.2$ eV, $\gamma_1 = 0.025$ eV, $\gamma_2 = 0$ eV, $v = 0.1$ eV$^2$. If for steady-state oscillations under the action of a harmonic force, the level of excitation can be traced by the amplitude of oscillations, for nonstationary modes it is more convenient to control the squared amplitude $\langle x_j^2(t) \rangle$ averaged over oscillation. In Fig. 1 we present the results of calculation for the time-dependent excitation of the first oscillator in the model system under the action of a monochromatic force in the vicinity of a Fano resonance during frequency scanning. The inset in Fig. 1 shows the frequency dependent response of the system on a monochromatic force demonstrating two characteristic resonance maxima. The second maximum near the frequency $\omega_2 = 1.2$ eV reflects a characteristic dip specific to a Fano resonance that means tendency the system response be zero.

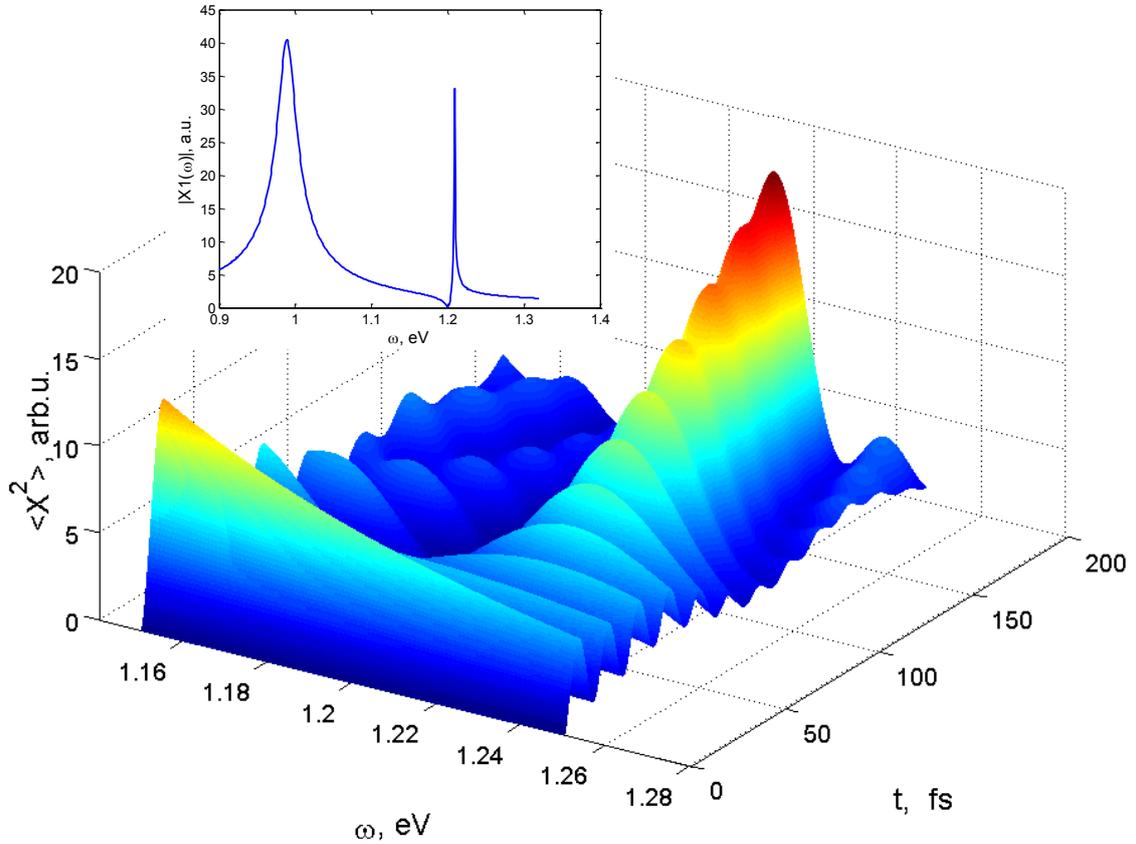

**Fig. 1.** The frequency-time dependence of a Fano resonance for two coupled oscillators excited by the harmonic force $f(t) = f_0 \sin \omega t$. The dispersion curve of reaction the first oscillator is presented as inset.

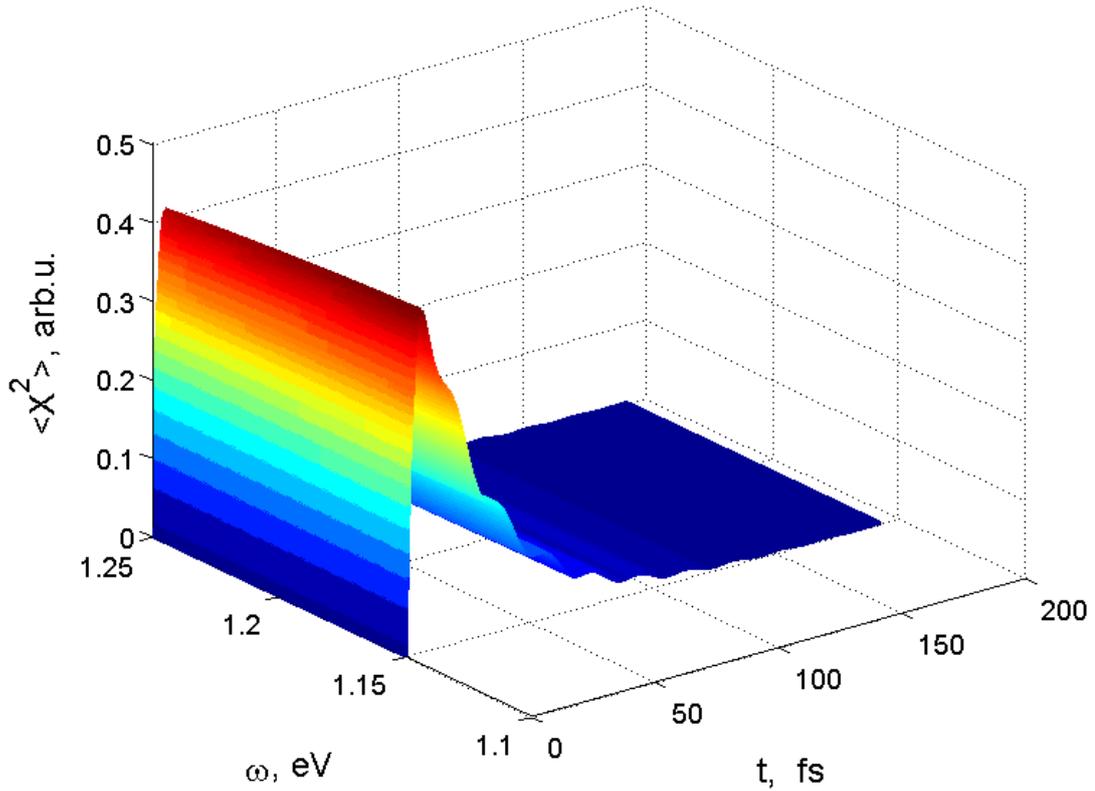

**Fig. 2.** The frequency-time dependence of a Fano resonance for a system of two coupled oscillators excited by the pulse $f(t) = f_0 e^{-t/\tau} \sin \omega t$, $\tau = 2.2$ fs.

For spectrally narrow resonances and a broadband external field, the details of the pulse excitation become structure inessential, the spectral distribution of a pulse remain steady inside a resonance range and the result is equivalent to the action of the $\delta$-pulse that is a convenient for representing of any ultrashort pulse. Fig. 2 shows the evolution of the model system excited by an ultra-broadband pulse. It illustrates independence the excitation dynamics with respect to the carrier frequency of an ultrashort pulse.

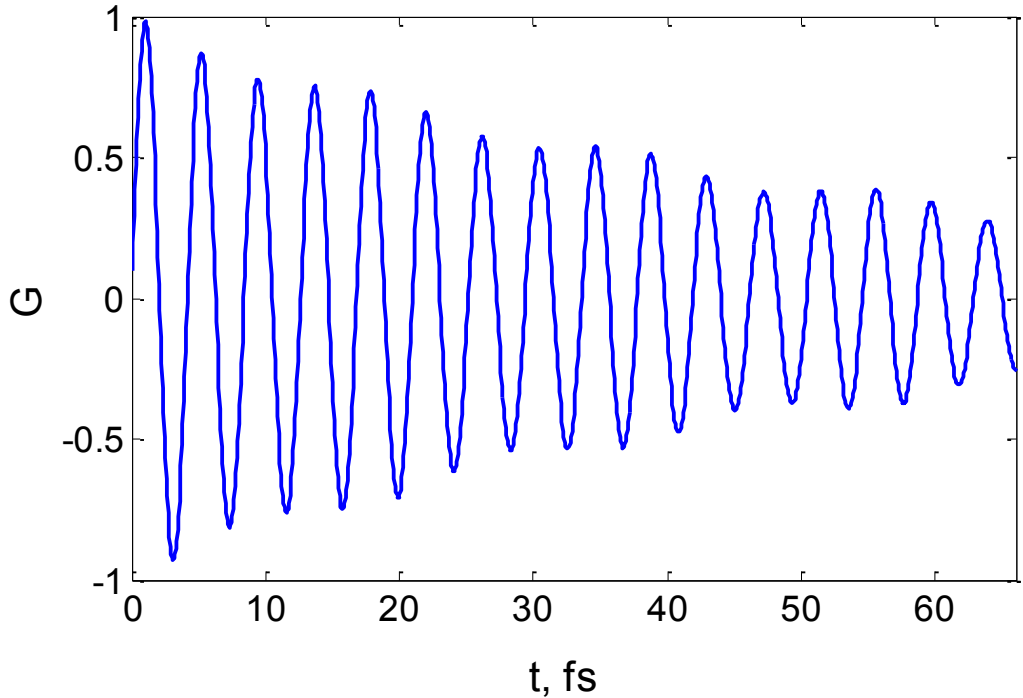

**Fig. 3.** The time-response function $G(t)$ of the model system.

In fact, the system dynamics is described by the solution of the system of equations (1) with $f(t) = \delta(t)$, being the corresponding response function $G(t)$. In the Laplace transform domain, such a pulse looks like $F(s) = 1$, and the response function in the time domain is a sum of damped oscillations satisfying the boundary conditions $x_1(0) = 0, \dot{x}_1(0) = 1$, $x_2(0) = 0, \dot{x}_2(0) = 0$. These conditions are physically means that the kick of the $\delta$-pulse changes momentum of the first oscillator by a unit amount without variation other boundary conditions of the system. The calculated response function is shown in Fig. 3 and is the sum of two damped vibrations with a beating time period 82.6 fs.

The model of coupled classical oscillators was lucky applied to describe the plasmon resonances in nanostructures [39-42]. For a system consisting of an optically coupled quantum dot (nanocrystal) and a metal nanoparticle [42], the parameters of a plasmon resonance were $\omega_1 = 2.118\,\text{eV}$, $\gamma_1 = 55.7\,\text{meV}$. The parameters of a quantum dot were $\omega_2 = 2.11\,\text{eV}$, $\gamma_2 = 2\,\text{meV}$. The coupling constant was $v = 60\,\text{meV}^2$. Since the model is linear, the incident field magnitude influences only the absolute value of response amplitude and has no effect on its frequency or time dependence, so the excitation amplitude is chosen arbitrarily. Fig. 4 shows the time-frequency dependence of the system excitation which tends to the experimental profile data [42] for large times.

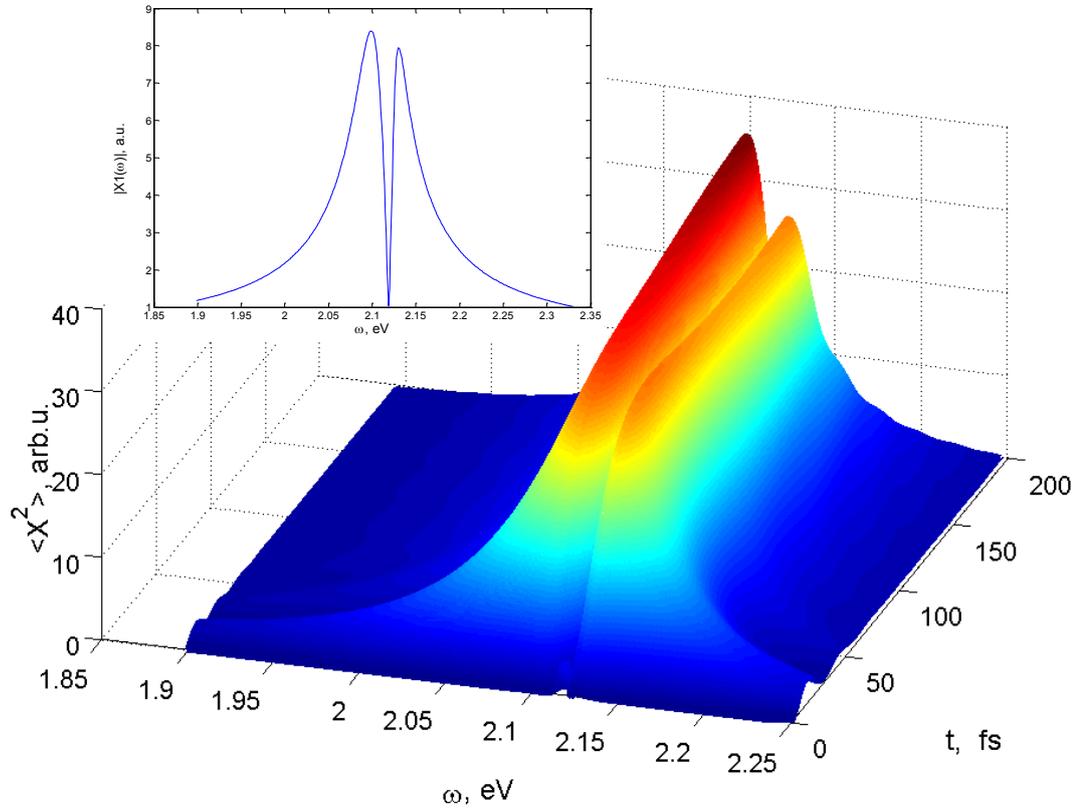

**Fig. 4.** The frequency-time dependence of a Fano resonance for an optically coupled quantum dot (nanocrystal) and a metal nanoparticle excited by the field $f(t) = f_0 \sin \omega t$. The dispersion dependence of the system is presented as inset.

It demonstrates the build-up of a Fano resonance under the action of a monochromatic field. After a certain transient process, a stationary pattern with a characteristic narrow dip at the spectral center is set.

### 3. Asymmetric resonances in coupled plasmonic systems

Now we are in a position to consider much more prominent example gives us a tunable Fano resonance nanostructure consisting of four interacting nanorods [43]. The Fano resonance in this structure is caused by interference of dipole modes. This system can be described by extended coupled oscillator (ECO) model. In ECO model the system of two oscillators is characterized by resonant frequencies $\omega_1, \omega_2$ and damping $\gamma_1, \gamma_2$, according to nonradiative losses. The radiative damping of two dipole oscillators is expressed in terms of the total dipole moment of the system as $P_{tot} = P_1 + P_2 = \alpha_1 x_1 + \alpha_2 x_2$, where $P_{1,2}$ are dipole moments of oscillators 1 and 2, $x_{1,2}$ are their amplitudes and $\alpha_{1,2}$ are polarizabilities. The spatial extension of the system is assumed to be smaller than a quarter wavelength of the incident light and the forces applied on both oscillators to be in phase. The net force is proportional to their polarizabilities $f_{1,2} = \alpha_{1,2} E_{ext}$.

The equation of motion can be written as follows:

$$\ddot{x}_1 + \gamma_1 \dot{x}_1 + \omega_1^2 x_1 + v x_2 = \frac{1}{2} \dddot{P}_{tot} + \alpha_1 E_{ext}(t), \tag{8}$$

$$\ddot{x}_2 + \gamma_2 \dot{x}_2 + \omega_2^2 x_2 + v x_1 = \frac{1}{2} \dddot{P}_{tot} + \alpha_2 E_{ext}(t).$$

The term $\frac{1}{2}\dddot{P}_{tot}$ introduces radiative coupling between oscillators. For an incident harmonic field $E_{ext}(t) = E_0 e^{i\omega t}$ the displacements $x_{1,2}$ are harmonic with $x_{1,2} = X_{1,2} e^{i\omega t}$, where oscillation amplitudes are

$$X_1(\omega) = \frac{\alpha_1 a_{22} - \alpha_2 a_{12}}{a_{11} a_{22} - a_{12} a_{21}} E_0, \tag{9}$$

$$X_2(\omega) = \frac{\alpha_2 a_{11} - \alpha_1 a_{21}}{a_{11} a_{22} - a_{12} a_{21}} E_0.$$

Here coefficients $a_{km}$ are defined as $a_{kk} = \omega_k^2 - \omega^2 + i\gamma_k \omega + \frac{i}{2}\alpha_k \omega^3$, $a_{12} = v + \frac{i}{2}\alpha_1 \omega^3$, $a_{21} = v + \frac{i}{2}\alpha_2 \omega^3$. $\omega_1 = 1.55\,\text{eV}$, $\omega_2 = 1.78\,\text{eV}$, $\gamma_1 = 0.083\,\text{eV}, \gamma_2 = 0.051\,\text{eV}$, $v = 0.084\,\text{eV}^2$ $\alpha_1 = 0.128\,\text{eV}^{-1}, \alpha_2 = 0.131\,\text{eV}^{-1}$ [43]. For arbitrary time dependence, the light scattering efficiency of the system is represented by squared modulus of the electric far field $|E_{sc}|^2 \sim |\ddot{x}_1 + \ddot{x}_2|^2$ averaged over oscillation. For an incident harmonic field it is proportional to the squared modulus of amplitudes $\omega^4 |X_1 + X_2|^2$ and can be used to model the scattering spectrum [44]. For close resonance frequencies $\omega_1$ and $\omega_2$ one can assume $\omega$ as having average value inside frequency interval. Consequently, for a previously considered system consisted of a quantum dot and metal nanoparticle with $(\omega_2/\omega_1)^4 = 1.015$, we can put this value a constant. However, for the present case of plasmonic nanoantenna it is not true, because resonances have rather different frequencies. This conclusion is confirmed by Fig. 5, where the difference between amplitude response and field scattering is quite clear. It means that parameterization, presented in [43], need further specification.

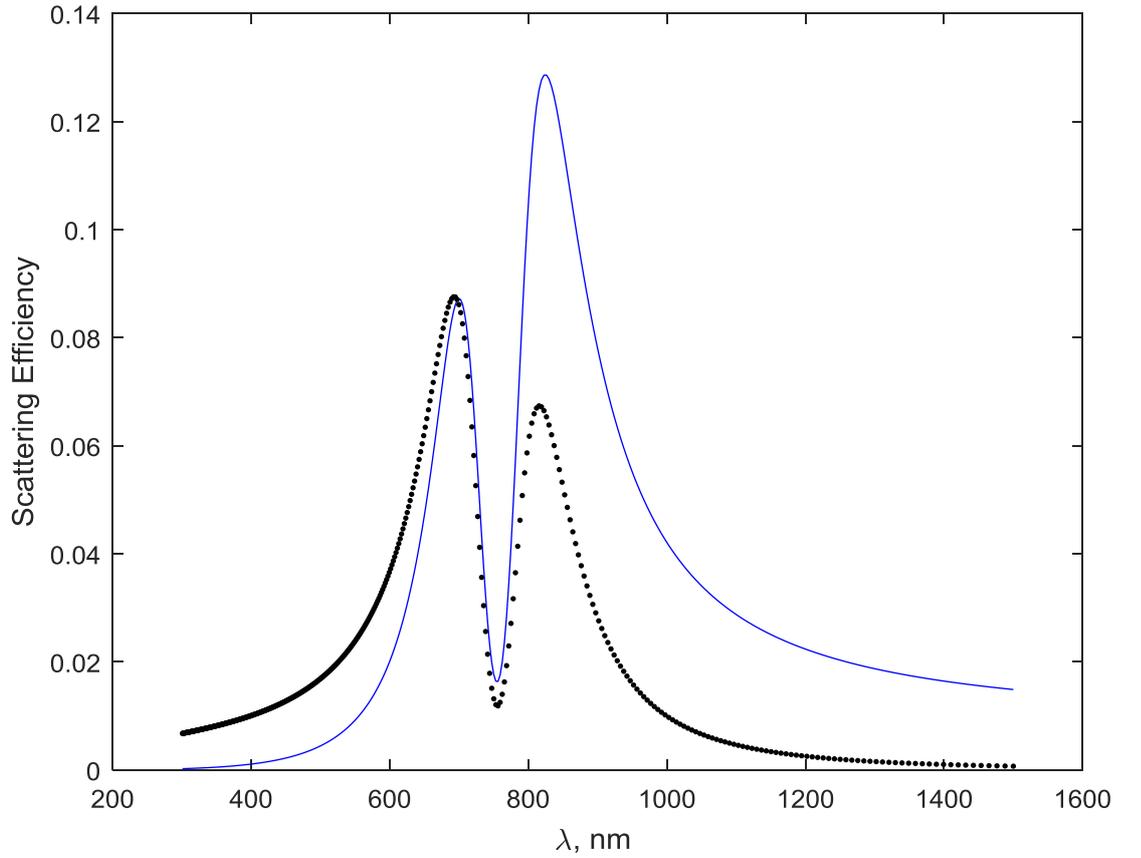

**Fig. 5.** Dipole frequency response of the system $|X_1 + X_2|^2$ (solid line) and scattering efficiency dependence of frequency (dotted line).

We fitted experimental scattering efficiency [43] taking into account actual factor $\omega^4$. The parameterization of the system, presented in [43], after more accurate specification are defined as $\omega_1 = 1.46$ eV, $\omega_2 = 1.64$ eV, $\gamma_1 = 0.010$ eV, $\gamma_2 = 0.195$ eV, $v = 0.14$ eV$^2$, $\alpha_1 = 0.195$ eV$^{-1}$, $\alpha_2 = 0.045$ eV$^{-1}$. Further calculations will be carried out with the specified parameters of the system, as well as taking into account $\omega^4 |X_1 + X_2|^2$ response. As shown in Fig. 6., the experimental and model frequency dependence are rather close at the vicinity near the maxima and minimum.

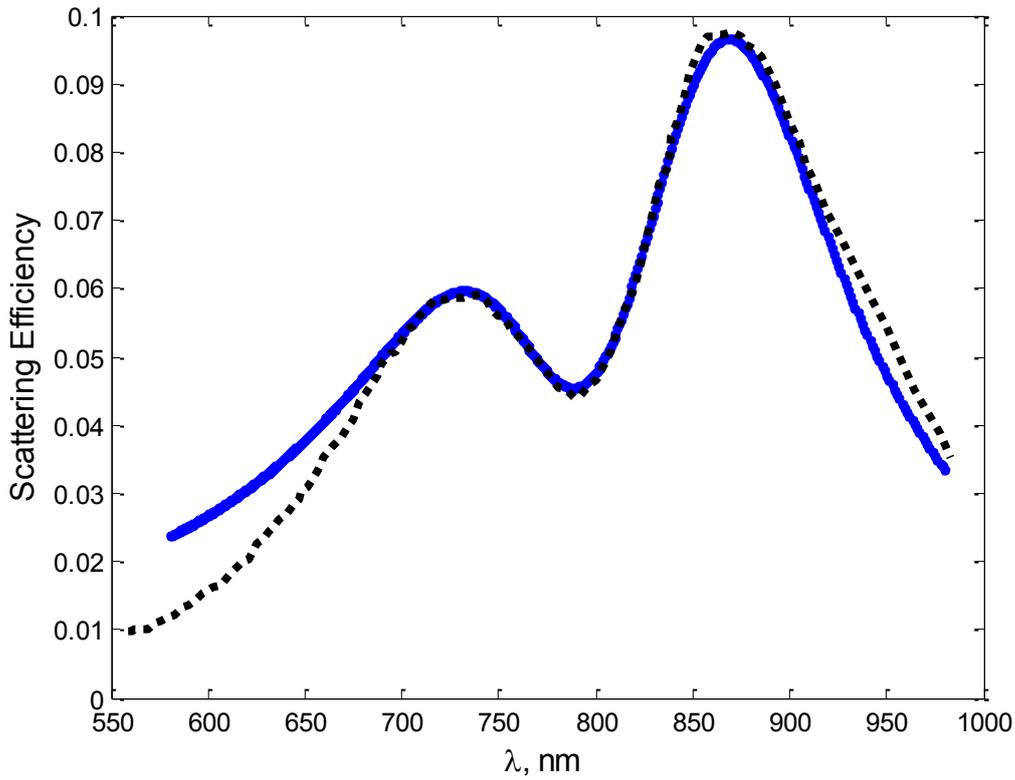

**Fig. 6.** Dipole frequency response of the system $\omega^4|X_1+X_2|^2$ (solid line) and experimental scattering efficiency dependence as a function of frequency [43] (dotted line).

The time-dependent response of the system on a pulse we obtain following the way described above. Fig. 7 shows the time response function of the system as a solution $G = x_1(t)+x_2(t)$ after terminating the incident ultrashort pulse having a $\delta$-pulse form and response in the form of acceleration $a(t)=\ddot{G}(t)$. Correlation coefficient $R = \mathrm{Corr}(G,a) = -0.937$ indicates some deviation from exact identity of functions (excluding sign minus), that would be valid for a response of a system on harmonic signal.

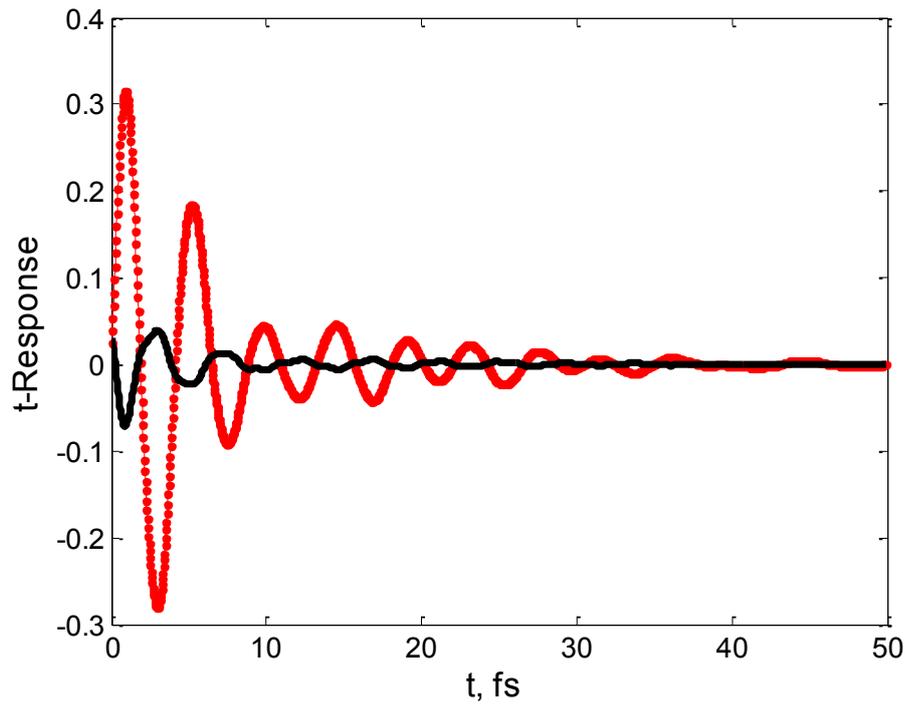

**Fig. 7.** Dipole response $G(t)$ (dotted line) and acceleration $a(t)$ (solid line) that is proportional to scattered light field.

The scattered field $E_{sc}(t)$ for the incident light with arbitrary time dependence $E_{ext}(t)$ is given by convolution

$$E_{sc}(t) = \int_{-\infty}^{t} a(t-t_1) E_{ext}(t_1)\, dt_1 . \qquad (10)$$

Taking into account the revealed characteristic ultrashort pulses in the optical range, a convenient form of their representation are the Gabor signals with a Gaussian envelope $E_{ext}(t) = e^{-(t/\tau)^2} \sin \omega t$. Fig. 8 shows the averaged over oscillation scattering intensity as a function of time and the incident pulse carrier frequency for a pulse with duration $\tau = 10$ fs. The two-hump structure of light scattering in a dynamical process of pulse scattering, embodied in the spectral distribution, is clearly manifested.

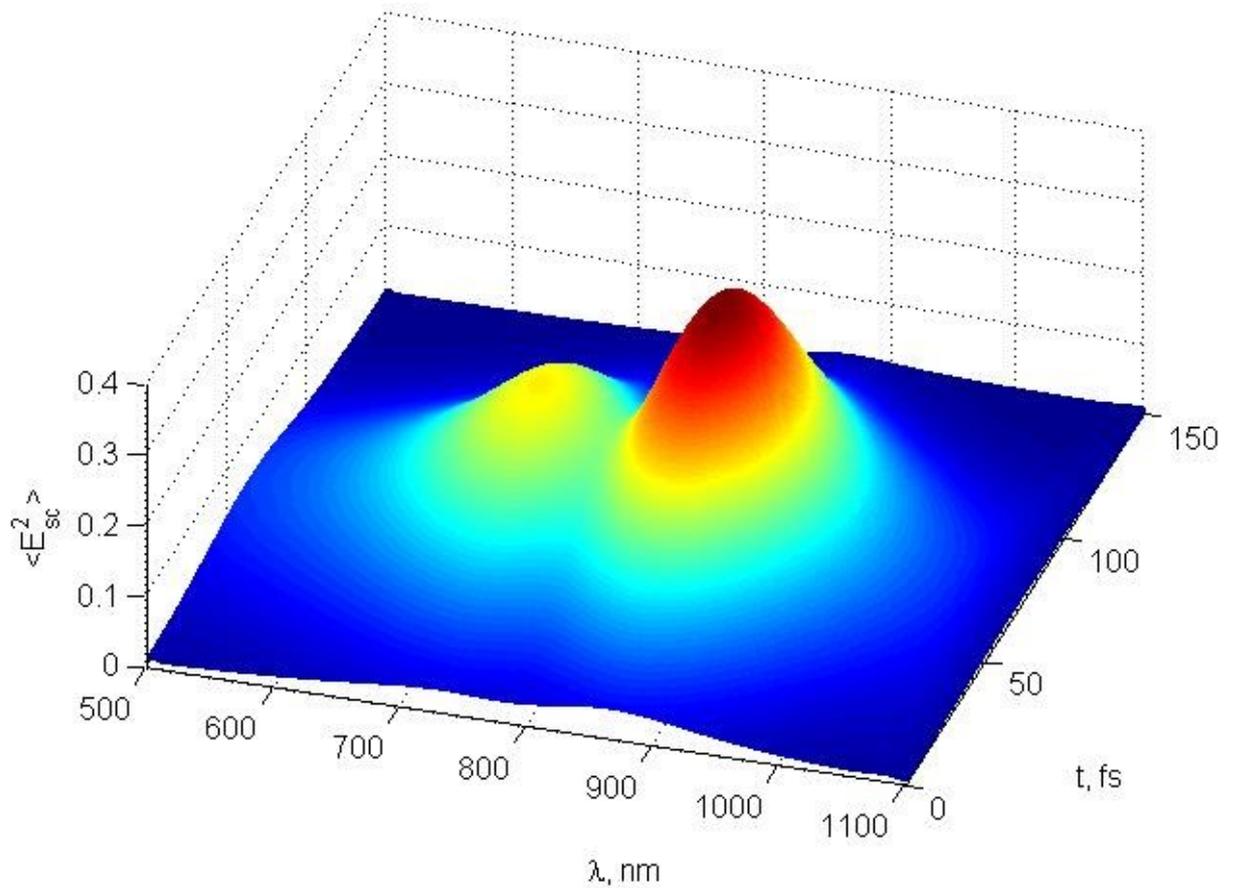

**Fig. 8.** Scattering intensity as a function of time and carrier wavelength of the incident pulse.

If we will measure the total spectrum of a scattered light, it can be calculated simply as a product the scattering efficiency in the frequency domain on the spectrum intensity of the incident pulse. The resulting spectrums are presented in Fig. 9. The solid line for relatively long pulse with duration 10 fs reproduces the incident light spectrum with the magnitude according scattering function. The shorter pulse with $\tau = 2.2$ fs demonstrate some two-humped internal structures attributed to the two-frequency resonant nanoantenna response.

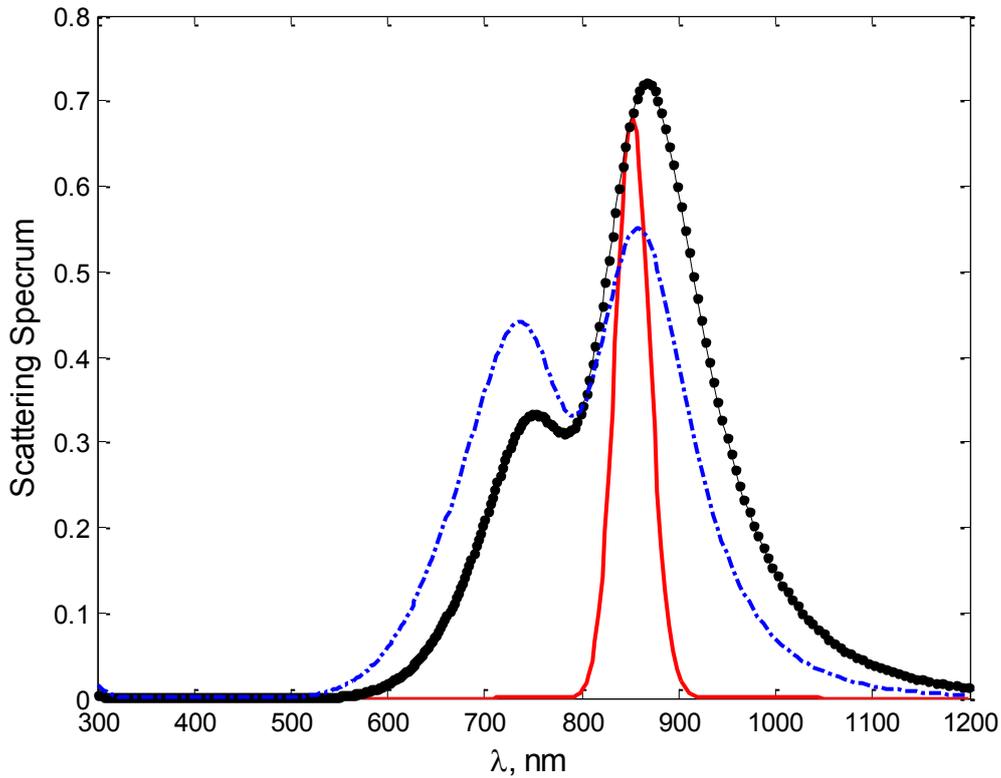

**Fig. 9.** Spectrum of the scattered Gabor pulses: solid line corresponds to $\tau = 10$ fs and the carrier pulse frequency $\omega = 1.55$ eV; for $\tau = 2.2$ fs the dotted line shows the result for $\omega = 1.55$ eV and the bar-dotted line corresponds to the carrier frequency 1.78 eV.

## 4. Conclusions

The results we have obtained on the basis of the model of coupled classical oscillators for the time dependence a Fano profile formation in fact have a more wide range of applicability since the linear response of arbitrary system is completely described by its dispersion properties. This conclusion follows from representation of linear response to a pulse of an arbitrary shape as the Fourier integral by including properly frequency dependence in a transfer function. Therefore the choice of specific realization of a linear operator having an adequate spectral response has no effect on the properties being described. This allows the use of the coupled classical oscillators model to describe the dynamics a great variety of systems with different resonance characteristics. However, the overall time response of a system significantly depends on the shape of an incident pulse.

The frequency characteristics of the response function can be measured experimentally by frequency scanning using spectrally narrow frequency-tunable radiation or by the technique of response restoration by measurement a correlation function for ultrashort pulses [45]. Another universal method to obtain the total characteristic of a linear system is the measurement of a response to the $\delta$-pulse. All extremely short pulses produce close, in fact universal, responses when their spectral width exceeds considerably the width of the system resonances. Hence, for different types of atomic, molecular and nanosized systems, femto- or attosecond laser pulses can be used as such probing pulses. It should be noted that the experimental demonstration of our general conclusions can be carried out, in particular, even with classical mechanical oscillatory systems, in RLC circuits as well as by measurement of light pulse scattering by metal subwavelength objects of different shapes [46].

**Acknowledgement**

The work was financially supported by the State Contract of the RF Ministry of Education and Science (assignment No. 3.9890.2017/8.9).